\documentclass[preprint]{aastex631}

\shorttitle{Continuous Habitable Zones}
\shortauthors{Ware et al.}
\graphicspath{{./}{figures/}}

\begin{document}

\title{Continuous Habitable Zones: Using Bayesian Methods to Prioritize Characterization of Potentially Habitable Worlds}

\author[0000-0003-2828-0334]{Austin Ware}
\affiliation{School of Earth and Space Exploration \\
Arizona State University \\
Tempe, AZ 85287, USA}

\author[0000-0003-1705-5991]{Patrick Young}
\affiliation{School of Earth and Space Exploration \\
Arizona State University \\
Tempe, AZ 85287, USA}

\author[0000-0002-1449-3008]{Amanda Truitt}
\affiliation{Los Alamos National Laboratory \\
Los Alamos, NM 87545, USA}

\author[0000-0003-3712-6769]{Alexander Spacek}
\affiliation{Los Alamos National Laboratory \\
Los Alamos, NM 87545, USA}



\begin{abstract}

The number of potentially habitable planets continues to increase, but we lack the time and resources to characterize all of them. With $\sim$30 known potentially habitable planets and an ever-growing number of candidate and confirmed planets, a robust statistical framework for prioritizing characterization of these planets is desirable. Using the $\sim$2 Gy it took life on Earth to make a detectable impact on the atmosphere as a benchmark, we use a Bayesian statistical method to determine the probability that a given radius around a star has been continuously habitable for 2 Gy. We perform this analysis on 9 potentially habitable exoplanets with planetary radii $<$1.8 R$_\oplus$ and/or planetary masses $<$10 M$_\oplus$ around 9 low-mass host stars ($\sim$0.5-1.1 M$_\odot$) with measured stellar mass and metallicity, as well as Venus, Earth, and Mars. Ages for the host stars are generated by the analysis. The technique is also used to provide age estimates for 2768 low-mass stars (0.5-1.3 M$_\odot$) in the TESS Continuous Viewing Zones.

\end{abstract}



\section{Introduction} \label{sec:intro}

The search for habitable worlds is at the forefront of astronomy and astrobiology. Over 4000 currently confirmed exoplanets and over 5000 TESS candidates  \citep{2014SPIE.9143E..20R} are currently cataloged. It is now known that exoplanets are fairly common in the habitable zones (HZ) of Sun-like and M-type stars \citep[e.g.][]{2007AsBio...7...30T,2011ApJ...729...27B,2013PNAS..11019273P,2015ApJ...807...45D,2020AJ....159..279B}, and rocky planets have been found inside HZs \citep[e.g.][]{2017Natur.544..333D,2017Natur.542..456G,2021AJ....161...36B}. Preliminary target lists have been mooted for NASA direct imaging missions: HabEx \citep{2020arXiv200106683G}, LUVOIR A/B \citep{2019arXiv191206219T}, and Starshade Rendezvous \citep{2019BAAS...51g.106S}. Therefore, it is now important to quantitatively characterize these known and potential planetary systems in terms of potential for habitability. The large number of theoretical and observational uncertainties mean that any given planet can only be assigned a probability that it is within the HZ and has been for some amount of time. With many known HZ planets, and additional HZ planets likely to be discovered by current and future missions, a robust statistical framework with estimates of as many contributing probability distributions is desirable.

Each system with planets inside the HZ is of interest due to the potential for the emergence of life. However, we assume that life's ability to establish itself relies largely on the stability of the planet's environment \citep[e.g.][]{2014PNAS..11112628M,2018PNAS..115..260D}. We therefore need to know how long a planet has resided within the HZ to accurately predict the likelihood of life emerging and making a detectable impact on the planet's environment. Determining the length of time a planet spends in the HZ is difficult due to the various uncertainties in the actual definition of the HZ and in the characterization of the planetary system. Using the time it took for life on Earth to significantly alter the atmosphere as a benchmark, we can use a Bayesian method to determine the likelihood that a planet in a given system will spend a similar amount of time in the continuous habitable zone (CHZ).

Previous work has shown that life on Earth made a detectable impact on the atmosphere after $\sim$1-2 Gy following the Earth's formation \citep{1993Icar..101..108K,Brocks1033,2005PNAS..10211131K,2007Sci...317.1903A,2013Natur.501..535C}, with the Great Oxidation Event (GOE) occurring around 2 Gy following formation \citep{1999Natur.400..554S,2003ARA&A..41..429K,doi:10.1098/rstb.2006.1838}. Following the work of \citet{2015ApJ...804..145T}, \citet{2017ApJ...835...87T}, and \citet{2020AJ....159...55T}, we therefore define the orbital area around a host star that will remain habitable for at least 2 Gy, here called the 2 Gy continuous habitable zone (CHZ$_2$), as a conservative estimate of the potential habitability of planets in the system across the entire main sequence. More importantly, for specific planets, it is necessary to determine whether they have already spent 2 Gy in the HZ. We acknowledge that different timescales could be adopted considering the HZ lifetimes of planets around various stellar types \citep{2013AsBio..13..833R}, but using Earth's evolutionary history is a useful starting point in order to narrow down the list of potentially habitable planets. Different timescales can also be adopted considering the history of disequilibrium chemistry in the Earth's atmosphere. As far back as the Archean eon (4.0-2.5 Ga), likely levels of biogenic CH$_4$ spurred detectable disequilibrium chemistry \citep{2018SciA....4.5747K}. However, the magnitude of the Earth's atmospheric disequilibrium increased with time in correlation with increased biomass and the evolution of oxygenic photosynthesis. Therefore, we find 2 Gy to be a sufficiently conservative amount of time to allow life to make a large, detectable impact on a planet's atmosphere. Our framework is flexible and allows this limit to easily be adjusted in the future.

Placing a likelihood on the current potential habitability of a detected planet will require knowing the age of the system, but difficulties in directly measuring stellar ages \citep{2010ARA&A..48..581S} means field stars typically lack age measurements. By fitting measurements of stellar properties to stellar evolutionary tracks, we can calculate approximate ages for field stars. Uncertainties on estimates of stellar ages for Sun-like stars are often on the order of 1 Gy \citep{2010ARA&A..48..581S,2010A&ARv..18...67T}. However, even a poorly known age can influence the prioritization of planets by considering the necessary time for life to produce detectable biosignatures in their atmospheres. Stellar age estimates then prove critical for stars without proper age measurements.

Along with our age estimates from fits to stellar evolution tracks, we incorporate age measurements from the literature. A common observational measurement method for field stars is gyrochronology, an approach to estimating the age of a low-mass star via its rotation period. Over the course of a typical solar-type star's life \citep[F7-K2 type,][]{2008ApJ...687.1264M}, magnetic breaking leads to the loss of angular momentum through mass loss \citep{1967ApJ...148..217W}, which in turns causes a slowdown in the rotation rate \citep{1972ApJ...171..565S}. By comparing field star rotation periods to stars with known ages and periods, the age of the star can be roughly constrained. \citet{2007ApJ...669.1167B} presents a color-dependent law for the rotational period of solar-type stars, which can be used to create \emph{gyrochrones} or plots of stars with the same age, but differing periods and colors. If a star's rotational period and color is known, they can be matched to the nearest gyrochrone and give an estimate of the age.

Samples of potentially habitable exoplanets can vary greatly depending on the HZ and terrestrial planet limits. Determining which planets around a star lie within the HZ requires using a model to determine where liquid surface water could potentially exist. These models can vary from a simple consideration of the stellar flux to more complex models considering a range planet masses and atmospheric compositions. Depending on the size of the estimated HZ, contrasting models can predict a significantly different sample of potentially habitable planets. Using optimistic constraints for the HZ \citep{2013ApJ...765..131K} and rocky-planet boundaries \citep{2016ApJ...819..127Z,2017AJ....154..109F,2018ApJS..235...38T}, we estimate there to be $\sim$30 currently known potentially habitable exoplanets.

With these goals in mind, we use a Bayesian statistical method to determine the likelihood that a given radius from each star has been continuously habitable for 2 Gy \citep{2020AJ....159...55T} for 9 potentially habitable exoplanets around 9 0.5-1.1 M$_\odot$ host stars with planetary radii $<$1.8 R$_\oplus$ and/or masses $<$10 M$_\oplus$ (see Section \ref{sec:samp}), as well as Venus, Earth, and Mars. The results are intended to inform target prioritization for future NASA missions, such as LUVOIR and HabEx, and ground-based follow-up aimed at characterizing potentially habitable worlds. For planets that are not viable candidates for follow-up with these missions due to distance or other factors, this procedure provides characterization that can be validated by the observable planets, providing a statistical estimate of the habitability potential of the local region of the Galaxy. The method also enables rapid characterization of future discoveries in the working angles of HabEx and LUVOIR target stars.

We use stellar mass, metallicity, and age, with their associated uncertainties, to be our observational priors. Because stellar ages are not known for many systems, we also develop an algorithm to determine best-fit model ages, generated using the Tycho database of stellar evolution models, for potentially habitable systems. We compare best-fit model ages for the sample of 9 potentially habitable exoplanet host stars to known ages and perform the CHZ$_2$ analysis. As an addition to this work, we determine best-fit ages for a sample of 2768 TESS Continuous Viewing Zone (CVZ) F, G, K, and early-M stars between 0.5-1.3 M$_\odot$. Future work includes expanding this Bayesian method to include additional stellar and planetary properties and testing various HZ model prescriptions.

\section{Methods}\label{sec:meth}

\subsection{Sample Selection}\label{sec:samp}

We chose a sample of 9 potentially habitable exoplanets around host stars approximately 0.5-1.1 M$_\odot$ with the lower limit determined by the current lower mass limit on the Tycho Stellar Evolution Catalog. Parameters for the sample of exoplanets are listed in Table \ref{tab:samp} and the associated host star parameters are listed in Table \ref{tab:hab_params}. The planets have radii $<$1.8 R$_\oplus$ and/or masses $<$10 M$_\oplus$, putting them within the optimistic range for rocky planets. The upper radius limit for rocky planets is adopted from \citet{2017AJ....154..109F,2018ApJS..235...38T}. 1.8 R$_\oplus$ is the approximate center of the 1.5-2.0 R$_\oplus$ gap between rocky ("super-Earth") and gaseous ("mini-Neptune") planets, indicating an optimistic threshold. For candidates with mass measurements, we adopt the upper mass limit derived from the planetary mass-radius relation for rocky cores from \citet{2016ApJ...819..127Z}, $M=R^{1.37}$. Assuming the above upper radius limit of 1.8 R$_\oplus$, we determine a value of $\sim$9 M$_\oplus$ and round up to 10 M$_\oplus$.

\begin{deluxetable}{ccccccc}[htb!]
    \tablecolumns{7}
    \tabletypesize{\scriptsize}
    \tablewidth{0pt}
    \tablecaption{Selected Potentially Habitable Planets\label{tab:samp}}
    \tablehead{\colhead{Planets} & \colhead{M$_\oplus$} & \colhead{R$_\oplus$} & \colhead{Period [d]} & \colhead{Ref.} & \colhead{NexSci Conf.?\tablenotemark{a}}}
    \startdata
    Kepler-1455 b & ... & 1.75 & 49.27684 & 1 & Y \\
    Kepler-438 b & ... & 1.12 & 35.23319 & 2 & Y \\
    KIC-7340288 b & ... & 1.51 & 142.5324 & 3 & N \\
    Kepler-441 b & ... & 1.462 & 207.2482 & 4(R),1(P) & Y \\
    Kepler-442 b & ... & 1.34 & 112.3053 & 2 & Y \\
    HD 40307 g & 7.1 & ... & 197.8 & 5 & Y \\
    Kepler-62 f & ... & 1.531 & 267.291 & 4(R),6(P) & Y \\
    Kepler-1544 b & ... & 1.685 & 168.8116 & 4(R),7(P) & Y \\
    Kepler-452 b & ... & 1.511 & 384.843 & 4(R),8(P) & Y \\
    \enddata
    \tablenotetext{}{References: 1.\citet{2018ApJS..235...38T} 2.\citet{2015ApJ...800...99T} 3.\citet{2020AJ....159..124K} 4.\citet{2018ApJ...866...99B} 5.\citet{2013AA...549A..48T} 6.\citet{2013Sci...340..587B} 7.\citet{2017AJ....154..264T} 8.\citet{2015AJ....150...56J}}
    \tablenotetext{a}{NexSci confirmed exoplanet via two different detection methods}
\end{deluxetable}

\begin{deluxetable}{ccccccccc}[htb!]
    \tablecolumns{9}
    \tabletypesize{\scriptsize}
    \tablewidth{0pt}
    \tablecaption{Stellar Parameters for Potentially Habitable Systems\label{tab:hab_params}}
    \tablehead{\colhead{Stars} & \colhead{$M/M_\odot$} & \colhead{$M$ Ref.} & \colhead{$T_{eff}$ [K]} & \colhead{$T_{eff}$ Ref.} & \colhead{$log(L/L_\odot$)} & \colhead{$L$ Ref.\tablenotemark{a}} & \colhead{[M/H]} & \colhead{[M/H] Ref.}}
    \startdata
    Kepler-1455 & 0.528$^{0.036}_{0.030}$ & 1 & 3899 $\pm$ 78 & 2 & -1.233$^{0.056}_{0.081}$ & 2(R,T) & -0.21 $\pm$ 0.11 & 3 \\[5pt]
    Kepler-438 & 0.544$^{0.028}_{0.043}$ & 1 & 3748 $\pm$ 112 & 4 & -1.357$^{0.142}_{0.138}$ & 4 & 0.16 $\pm$ 0.14 & 4 \\[5pt]
    KIC-7340288 & 0.57$^{0.02}_{0.01}$ & 5 & 3949$^{79}_{52}$ & 5 & -1.1902$^{0.0883}_{0.1110}$ & 6 & -0.31 $\pm$ 0.14 & 7 \\[5pt]
    Kepler-441 & 0.573 $\pm$ 0.026 & 1 & 4340 $\pm$ 87 & 8 & -1.067$^{0.048}_{0.053}$ & 8(R,T) & -0.58 $\pm$ 0.15 & 1 \\[5pt]
    Kepler-442 & 0.613 $\pm$ 0.03 & 1 & 4402 $\pm$ 88 & 8 & -0.862$^{0.048}_{0.053}$ & 8(R,T) & -0.37 $\pm$ 0.1 & 4 \\[5pt]
    HD 40307 & 0.71 $\pm$ 0.02 & 9 & 4827 $\pm$ 44 & 10 & -0.642$^{0.011}_{0.012}$ & 11 & -0.25 $\pm$ 0.029 & 10 \\[5pt]
    Kepler-62 & 0.727$^{0.029}_{0.059}$ & 1 & 4859 $\pm$ 97 & 8 & -0.595$^{0.048}_{0.052}$ & 8(R,T) & -0.37 $\pm$ 0.04 & 12 \\[5pt]
    Kepler-1544 & 0.743$^{0.034}_{0.030}$ & 3 & 4852 $\pm$ 97 & 8 & -0.604$^{0.048}_{0.052}$ & 8(R,T) & -0.08 $\pm$ 0.1 & 3 \\[5pt]
    Kepler-452 & 1.07$^{0.06}_{0.04}$ & 13 & 5772$^{63}_{65}$ & 13 & 0.089$^{0.062}_{0.067}$ & 8(R),13(T) & 0.23 $\pm$ 0.04 & 13 \\[5pt]
    \enddata
    \tablenotetext{}{References: 1.\citet{2017ApJS..229...30M} 2.\citet{2018ApJS..235...38T} 3.\citet{2017AJ....154..264T} 4.\citet{2015ApJ...800...99T} 5.\citet{2020AJ....159..124K} 6.\citet{2019AJ....158..138S} 7.\citet{2016MNRAS.457.2877G} 8.\citet{2018ApJ...866...99B} 9.\citet{2016AA...585A...5B} 10.\citet{2005ApJS..159..141V} 11.\citet{2008AA...487..373S} 12.\citet{2013Sci...340..587B} 13.\citet{2017AJ....154..108J}}
    \tablenotetext{a}{(R,T) indicates value calculated using referenced radius and effective temperature.}
\end{deluxetable}

All 9 planets have instellation values within the optimistic Recent Venus (RV) inner habitable zone (IHZ) and Early Mars (EM) outer habitable zone (OHZ) boundaries for their system, calculated using the equations defined in \citet{2013ApJ...765..131K} and discussed in Section \ref{sec:hz_mod}. From a HZ perspective, these are high-priority candidates for spectral characterization.

\subsection{Tycho Stellar Evolution Catalog}\label{sec:tycho}

We use the stellar evolution code Tycho \citep{2005ApJ...618..908Y} to expand the catalog of evolutionary tracks in \citet{2015ApJ...804..145T} and \citet{2017ApJ...835...87T}. Tycho is a 1D stellar evolution code that utilizes a hydrodynamic formulation of the stellar evolution equations. Tycho contains OPAL opacities \citep{1994ApJ...437..879A,1996ApJ...464..943I,2002ApJ...576.1064R}, utilizes a combined OPAL and Timmes equation of state \citep{1999ApJS..125..277T,2002ApJ...576.1064R}, gravitationally induced diffusion \citep{1994ApJ...421..828T}, general relativistic gravity, automatic rezoning, and an adaptable nuclear reaction network paired with a sparse solver. Low-temperature ($\sim$2400 K) opacities, which include dust grain opacities, were added in \citet{2017ApJ...835...87T} and are based on \citet{2005ApJ...623..585F} and \citet{2009ApJ...705L.123S}. Tycho uses an adaptable 177 element network up to $^{74}$Ge that is utilized throughout the evolution. The network uses REACLIB rates \citep{1999NuPhA.656....3A,2000ADNDT..75....1R,2001ApJS..134..151I,2006MmSAI..77..910W}, weak rates from \citet{2000NuPhA.673..481L}, and screening from \citet{1973ApJ...181..457G}. Mass loss is included, but is trivial for the mass range considered in this work. Neutrino cooling due to the Urca process and plasma processes is included. Turbulent convection is defined via a hydrodynamic formulation \citep{2007ApJ...667..448M,2009ApJ...690.1715A,2010ApJ...710.1619A,2011ApJ...733...78A} based on 3D, well-resolved simulations of convection between stable layers. Unlike stellar evolution codes relying on mixing-length theory, Tycho has no free convective parameters (i.e. "convective overshoot").

The catalog currently contains models between 0.5-1.3 M$_\odot$, with metallicities that fall between 0.1-3.0 of Z$_\odot$. The metallicity models are in steps of 0.1 Z$_\odot$ between 0.1-1.5 Z$_\odot$ and steps of 0.25 Z$_\odot$ between 1.5-3.0 Z$_\odot$. We added a finer grid of mass models, in steps of 0.05 M$_\odot$, between 0.5-1.0 M$_\odot$. The mass models are in steps of 0.1 M$_\odot$ between 1.0-1.3 M$_\odot$. We use these models in Section \ref{sec:hz_mod} to determine the HZ boundaries over the evolution of the star. The catalog also varies in [O/Fe] between 0.44x, 1.0x, and 2.28x [O/Fe]$_\odot$ for the full metallicity range. These models are included in Section \ref{sec:ages} when fitting the ages for stars.

While future direct detection mission such as HabEx and LUVOIR will concentrate on Sun-like stars, virtually all of the simulated HZ planet detections for TESS are around M-type and late K-type stars \citep{2015ApJ...809...77S,2018ApJS..239....2B}. There is then significant value in including host stars below 0.5 M$_\odot$. We will expand this catalog further and include stars down to 0.1 M$_\odot$ in a future paper.

\subsection{Habitable Zone Models}\label{sec:hz_mod}

Tycho outputs information on stellar surface quantities for each time-step of the evolution. For this work, we use the stellar effective temperature and luminosity to define the inner and outer boundaries of the HZ, as a function of stellar age, utilizing equations from \citet{2013ApJ...765..131K,2014ApJ...787L..29K}. This method can be used substituting other HZ prescriptions, but Kopparau et al. are a commonly used point of reference. These HZ prescriptions parameterize the location of the HZ as a function of the host star luminosity and effective temperature, from which we can calculate the associated time-dependent HZ distance for each stellar evolution track. For a given orbital distance from any star we can predict how long and at what stellar age a planet would remain habitable. Thus, for a perfectly characterized star (mass, metallicity, and age known exactly), we can say with high confidence whether a planet has been continuously within the circumstellar HZ for 2 Gy for a given HZ model. 

\citet{2013ApJ...765..131K,2014ApJ...787L..29K} give several possible definitions for the HZ boundaries. The most optimistic HZ definition follows from \citet{2013ApJ...765..131K}, where they empirically determine the inner and outer HZ edge assuming that Venus and Mars once hosted habitable conditions. For the inner habitable zone (IHZ) edge, the "recent Venus" (RV) case, they determined the effective solar flux on Venus 1 Gy ago, under the assumption that there may have been liquid water on the surface prior to this time. Similarly, for the outer habitable zone (OHZ) edge, the "early Mars" (EM) case, they determined the effective solar flux on Mars $\sim$3.8 Gy ago, when liquid water likely existed on the surface.

The conservative HZ definitions follow from \citet{2014ApJ...787L..29K}, where they define an IHZ edge by the "runaway greenhouse" case. Here, the effective solar flux incident on the planet becomes sufficient to completely vaporize the oceans. The OHZ edge is defined by the "maximum greenhouse" case. This outer edge is the point where Rayleigh scattering becomes dominant over the greenhouse effect of CO$_2$.  The conservative cases were calculated for masses of 0.1, 1, and 5 M$_\oplus$ to account for gravitational effects on the atmosphere.

\subsection{Bayesian Habitable Zone Probabilities}\label{sec:bayesian}

Bayesian statistics have been previously used to better understand the emergence of life on planets \citep{2012PNAS..109..395S} and statistical analysis has been used in \citet{2017ApJ...841L..24B} to develop a method for surveying key terrestrial exoplanet atmosphere composition characteristics, like H$_2$O and CO$_2$ abundances, in order to broadly assess many exoplanet habitability potentials, rather than spending time gathering extensive data about individual exoplanets. However, these methods focus on the characteristics of the exoplanet, but do not take into account the star and the stellar environment. \citet{2015ApJ...804..145T,2017ApJ...835...87T} discuss how the initial mass and composition will affect the surface properties, and therefore the HZ evolution \citep[e.g.][]{2013AsBio..13..833R,2017AsBio..17...61W}.

This Bayesian approach to HZs aims to determine the probability that a given orbital distance from the host star has spent 2 Gy in the HZ, following the methods of \citet{2020AJ....159...55T}. Depending on the known properties of the star, this approach could follow several different cases. The simplest example would involve knowing the mass and metallicity of the star arbitrarily well, but not knowing the age. The probability in this case would depend solely on how long the orbit remains in the HZ of the star as predicted by the models. In other cases, there is an additional or multiple uncertainties. If the metallicity and/or mass is unknown, then all metallicity and/or mass models must be integrated over. If the metallicity or mass is known to within some uncertainty, assumed to be Gaussian, then we must weight the contribution of each model by the fit of the model mass and model metallicity to the Gaussian of each measured value. Introducing the age for the star adds an additional Gaussian prior distribution to the calculation. Here, we focus on the cases where the metallicity and mass are known to within some uncertainty, but the age is unknown, as well as the case where the metallicity, mass, and age are known to within some uncertainty. The other cases are described in detail in \citet{2020AJ....159...55T}. Although this work is limited to considering metallicity, mass, and age, it can in principle be extended to include other properties, such as planetary composition and stellar activity.

\subsubsection{Case 1: Metallicity and Mass Measurement but No Age Measurement}\label{sec:case1}

We first describe the case where the stellar metallicity and mass are known to within some uncertainty, but the age is unknown. In this case, the age is limited to 12 Gy to account for the age of the Universe. Our method relies on the expansion of Bayes' Theorem:

\begin{equation}
    \centering
        P(A|B) = \frac{P(B|A) \times P(A)}{P(B|A) \times P(A) + P(B|\lnot A) \times P(\lnot A)}
\label{eq:1}
\end{equation}

\noindent where $P(A|B)$ is the posterior likelihood, or the likelihood of outcome $A$ occurring given $B$, $P(B|A)$ is the likelihood of $B$ given that $A$ is true, and $P(A)$ is the prior probability that the outcome $A$ is true.

We apply Equation (\ref{eq:1}) to the Tycho models and measured distributions for the stellar metallicity ($Z$) and mass ($M$) to calculate the Bayesian posterior probability. In Equation (\ref{eq:1}), $B=Z,M$ and $A=CHZ_2$, or the outcome where a given radius is in the CHZ$_2$. Equation (\ref{eq:1}) therefore becomes

\begin{equation}
    \centering
        P(CHZ_2|Z,M) = \frac{P(Z,M|CHZ_2) \times P(CHZ_2)}{P(Z,M|CHZ_2) \times P(CHZ_2) + P(Z,M|\lnot CHZ_2) \times P(\lnot CHZ_2)}.
\label{eq:2}
\end{equation}

We now describe how we compute each component of Equation (\ref{eq:2}) for a chosen model metallicity $Z_k$ and mass $M_k$, for any given radius from the star. The index $k$ refers to the specific model used, interpolated if necessary. We calculate $P(CHZ_2)$, the initial probability that a given radius is within the CHZ$_2$, by

\begin{equation}
    \centering
        P(CHZ_2) = \frac{\sum_i (\sum_j t_{CHZ_2,ij}\Delta Z_j)\Delta M_i}{\sum_i(\sum_j t_{tot,ij}\Delta Z_j)\Delta M_i}
\label{eq:3}
\end{equation}

\noindent where the index $i$ runs through the Tycho model masses $M = 0.5-1.3$ $M_\odot$ and the index $j$ runs through the model metallicities from $Z = 0.1-3.0$ $Z_\odot$, $t_{CHZ_2,ij}$ is the total time the radius is in the CHZ$_2$ for a given Tycho model ($M_i$,$Z_j$), $t_{tot,ij}$ is the total lifetime for the given Tycho model (with a maximum of 12 Gy), $\Delta M_i$ ($M_{i+1}-M_i$) is the distance between $M$ values, and $\Delta Z_j$ ($Z_{j+1}-Z_j$) is the distance between $Z$ values. Note that we first sum over all model metallicities and then over all model masses. The initial probability that a radius is not in the CHZ$_2$, P($\lnot CHZ_2$), is given by

\begin{equation}
    \centering
        P(\lnot CHZ_2) = 1-P(CHZ_2).
\label{eq:4}
\end{equation}

\noindent The likelihood that a star has a model metallicity $Z_k$ and mass $M_k$ if a given radius is in the CHZ$_2$ is given by

\begin{equation}
    \centering
        P(Z_k,M_k|CHZ_2) = \frac{t_{CHZ_2,k} \times P(Z_k) \times P(M_k)}{\sum_i(\sum_j t_{CHZ_2,ij} \times P(Z_j)) \times P(M_i)}
\label{eq:5}
\end{equation}

\noindent where $P(Z)$ and $P(M)$ are probabilities given by the Gaussian distributions for each measured value ($Z' \pm \sigma_Z$, $M' \pm \sigma_M$). For example, the measured metallicity distribution is given by

\begin{equation}
    \centering
        P(Z) = \frac{1}{\sqrt{2\pi}} e^{\frac{1}{2}\frac{\Delta 'Z^2}{\sigma ^2_Z}}
\label{eq:6}
\end{equation}

\noindent where $\Delta 'Z$ is defined as $Z'-Z$, the difference between the measured mean and model values. The likelihood that a star has metallicity $Z_k$ and mass $M_k$ if a given radius is not in the CHZ$_2$ is given by

\begin{equation}
    \centering
        P(Z_k,M_k|\lnot CHZ_2) = \frac{(t_{tot,k}-t_{CHZ_2,k}) \times P(Z_k) \times P(M_k)}{\sum_i (\sum_j (t_{tot,ij}-t_{CHZ_2,ij}) \times P(Z_j)) \times P(M_i)}.
\label{eq:7}
\end{equation}

We combine Equations (\ref{eq:3})-(\ref{eq:5}) and (\ref{eq:7}) to calculate the Bayesian posterior probability in Equation (\ref{eq:2}), $P(CHZ_2|Z_k,M_k)$. With a factor of $P(Z_k)$ and $P(M_k)$ cancelling out, we get

\begin{equation}
    \centering
        P(CHZ_2|Z_k,M_k) = \frac{\frac{t_{CHZ_2,k}}{\sum_i (\sum_j t_{CHZ_2,ij} P(Z_j)) P(M_i)} \frac{\sum_i \sum_j t_{CHZ_2,ij}}{\sum_i \sum_j t_{tot,ij}}}{\frac{t_{CHZ_2,k}}{\sum_i (\sum_j t_{CHZ_2,ij} P(Z_j)) P(M_i)} \frac{\sum_i \sum_j t_{CHZ_2,ij}}{\sum_i \sum_j t_{tot,ij}} + \frac{t_{tot,k}-t_{CHZ_2,k}}{\sum_i (\sum_j (t_{tot,ij}-t_{CHZ_2,ij}) P(Z_j)) P(M_i)} (1-\frac{\sum_i \sum_j t_{CHZ_2,ij}}{\sum_i \sum_j t_{tot,ij}})}.
\label{eq:8}
\end{equation}

We can now calculate the Bayesian posterior probability at each radius, for a given $Z$ and $M$, that it is in the CHZ$_2$ at any time during the main sequence lifetime, limited to 12 Gy. Therefore, probabilities cannot exceed 10/12 because each radii from the star must be habitable for at least 2 Gy, limiting the maximum CHZ$_2$ time to 10 Gy.

\subsubsection{Case 2: Metallicity, Mass, and Age Measurement}\label{sec:case2}

Without knowledge of the age, we previously summed the total CHZ$_2$ time and main sequence lifetime. If we know the age measurement in the form of Gaussian errors, $A \pm \sigma_A$, we can similarly use this as the probability term $P(A)$ using the Gaussian probability

\begin{equation}
    \centering
        P(A) = \frac{1}{\sqrt{2\pi}} e^{\frac{1}{2}\frac{\Delta 'A^2}{\sigma ^2_A}}
\label{eq:9}
\end{equation}

\noindent where $\Delta 'A$ is defined as $A'-A$. $P(A)$ is used to place a prior probability on the total and CHZ$_2$ times, $t_{tot}$ and $t_{CHZ_2}$, thereby prioritizing model timesteps closer to the mean age of the star in a similar way to how $P(Z)$ and $P(M)$ prioritize model metallicities and masses closer to the mean stellar metallicity and mass. $t_{tot}$ is now given by

\begin{equation}
    \centering
        t_{tot} = \sum_m(t_{tot,m}-t_{tot,m-1})P(A)
\label{eq:10}
\end{equation}

\noindent where the index $m$ runs through each model step and $A=t_{tot,m}$. By combining Equations (\ref{eq:10}) and (\ref{eq:8}), we can now calculate the Bayesian posterior probability that each radius is currently within the CHZ$_2$. We apply this method to our sample of potentially habitable planets, with results summarized in Table \ref{tab:prob} for each planet and the Bayesian posterior distributions for each star shown in Figure \ref{fig:res1}. For all 9 sample stars, we applied a 4x linear and cubic interpolation to the mass models.

\begin{figure}[htb!]
        \centering
        \begin{tabular}{@{}c@{}c@{}}
        \includegraphics[width=0.47\textwidth]{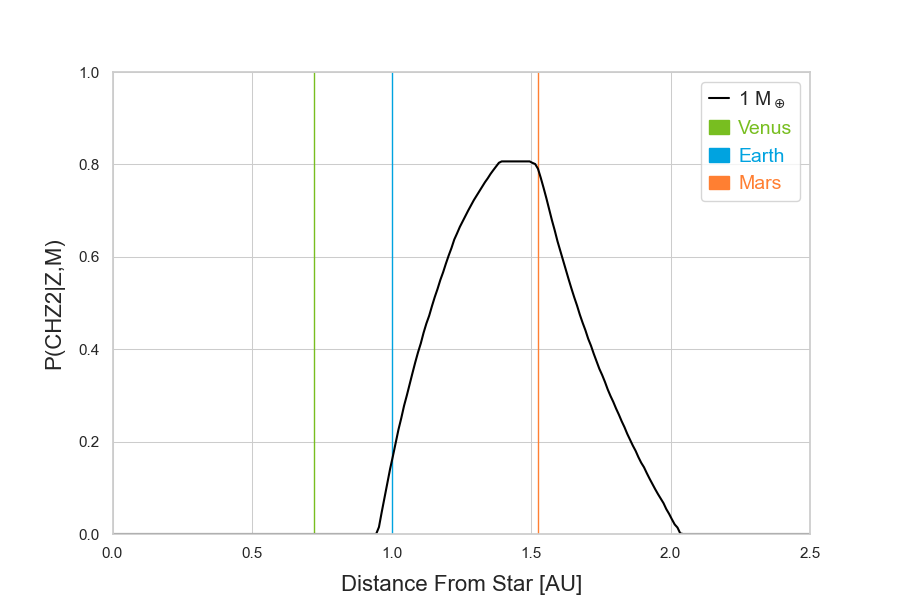} &
        \includegraphics[width=0.47\textwidth]{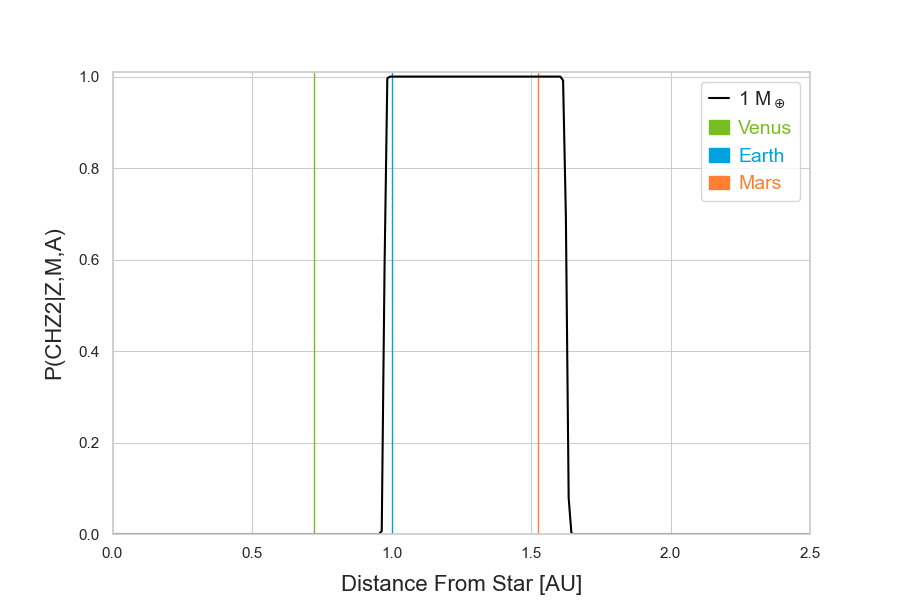}
        \end{tabular}
        \caption{P(CHZ$_2\mid$ Z,M) (left) and P(CHZ$_2\mid$ Z,M,A) (right) for the Sun assuming a 1 M$_\oplus$ planet with a runaway greenhouse conservative IHZ and maximum greenhouse conservative OHZ.}
        \label{fig:1}
\end{figure}

We use the methods for Case 1 and Case 2, as well as the $1 M_\oplus$ HZ model from \citet{2014ApJ...787L..29K}, to calculate the the CHZ$_2$ posterior likelihood for the Sun with and without the age prior. The age of the Sun, determined via helioseismology and solar models, is taken to be 4.57$\pm$0.11 Gy \citep{2002A&A...390.1115B}. The results are shown in Figure \ref{fig:1}, which includes the orbits of Venus, Earth, and Mars. This comparison exemplifies the extent to which a precise age measurement can have on the CHZ$_2$ probability distribution. Without an age measurement, Earth is given only a $\sim 20\%$ probability of being in the CHZ$_2$, while Mars has a $\sim 80\%$ probability. Earth's closer proximity to the Sun means it will leave the HZ sooner than Mars, significantly reducing the probability that it would currently be within the CHZ$_2$ without knowing the age. With knowledge of the Sun's age, we see that both Earth and Mars have a $100\%$ probability of being currently in the CHZ$_2$, which matches our current understanding of the Sun's HZ.

\subsection{Stellar Ages}\label{sec:ages}

Although most stars in this sample have measured ages, the vast majority of TESS targets and some known potentially habitable systems lack age measurements. Without measurements of the ages, we cannot determine how long planets have resided in the HZ. Therefore, it is essential to estimate the stellar age in order to more accurately predict the potential for life not only existing, but having made a detectable impact. Measurement of stellar ages is relatively straightforward if the star resides in a cluster \citep[e.g.][]{1970ApJ...162..841S,2002A&A...396..551L}, but is very difficult for field stars \citep{2010ARA&A..48..581S}. This often relies on estimates given by fits to stellar evolution models like Tycho. By using luminosity ($L$) and effective temperature ($T_{eff}$) measurements, we can perform $\chi^2$ fits to models from the Tycho stellar evolution catalog \citep{2015ApJ...804..145T,2017ApJ...835...87T} to determine the best-fit model ages:

\begin{equation}
    \chi^2 = \frac{(L_{mod}-L_{obs})^2}{\sigma^{2}_{L}}+\frac{(T_{mod}-T_{obs})^2}{\sigma^{2}_{T}}
\end{equation}

\noindent where $L_{mod}$ and $L_{obs}$ are the model and observed luminosities, $T_{mod}$ and $T_{obs}$ are the model and observed stellar effective temperatures, and $\sigma_{L}$ and $\sigma_{T}$ are the errors in the observations \citep{2001ApJ...556..230Y,2015ApJ...803...90P}. If mass and compositional measurements are available, the search is constrained to models bracketing the measured values of mass and metallicity. Model points were averaged and weighted by their associated $\chi^2$ value. Upper and lower uncertainties were derived from the weighted standard deviation of the sample.

Uncertainties on estimates of stellar age for Sun-like stars are rarely less than 1 Gy, often times even significantly larger \citep{2010ARA&A..48..581S,2010A&ARv..18...67T}. By operating within a Bayesian statistical framework, we can still extract useful information from a roughly constrained age. Assuming a Gaussian distribution for the best-fit age and uncertainty, we introduce the age as a prior probability distribution in the calculation. Even a poorly known age can then influence the prioritization of planets by taking into account that some planets may not have had enough time for life to produce detectable biosignatures in their atmospheres. The best-fit ages will then prove useful to determining the current habitability of the planets in these systems.

\begin{figure}[htb!]
    \centering
    \includegraphics[width=0.65\textwidth]{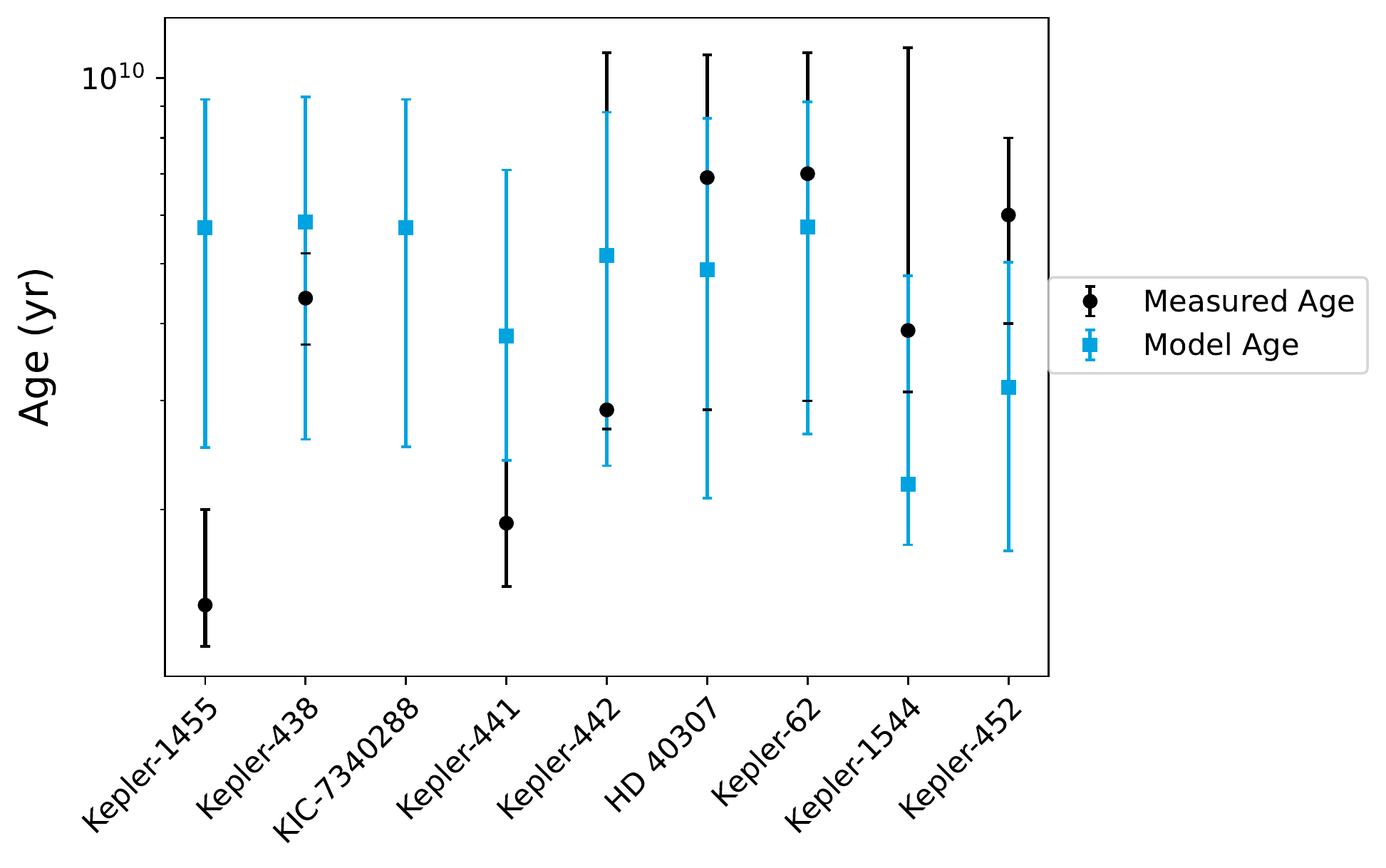}
    \caption{Comparison of measured literature ages and Tycho stellar evolution catalog best-fit model ages for sample of potentially habitable systems.}
    \label{fig:age_comp}
\end{figure}

Ages for 8 stars, except KIC-7340288, were available in the literature. We also generated best-fit ages for all of the stars in the sample. The ages for all 9 host stars along with source references are provided in Table \ref{tab:hab_ages}. We prefer stellar ages determined via gyrochronology for use as the age prior in \S \ref{sec:chz2_prof} to determine the P(CHZ$_2$) profiles for each system. Ages determined via gyrochronology tend to have tighter constraints on the age and the method is observationally calibrated, rather than relying on fits to isochrones or evolutionary tracks. For those stars with only isochrone ages, we take the average of the literature age and our fits to evolutionary tracks. For Kepler-442, which has a large upper error of $\sim 8$ Gyr, we also average the literature age with our age. Figure \ref{fig:age_comp} shows an over-plot of the measured literature ages and the best-fit model ages. Aside from Kepler-1455 and Kepler-441, the observed values fall within the predicted range of values determined by the models. Kepler-1455 and Kepler-441 are near the lower mass tail of the model space ($\sim$0.5 M$_\odot$) and will likely benefit from further increasing the mass and metallicity resolution and range of the model space.

\begin{deluxetable}{ccccc}[htb!]
    \tablecolumns{5}
    \tabletypesize{\scriptsize}
    \tablewidth{0pt}
    \tablecaption{Stellar Ages for Potentially Habitable Systems\label{tab:hab_ages}}
    \tablehead{\colhead{Stars} & \colhead{Age$_{Tycho}$ [Gy]} & \colhead{Age$_{lit}$ [Gy]} & \colhead{Ref.\tablenotemark{a}} & \colhead{Tech.\tablenotemark{b}}}
    \startdata
    Kepler-1455 & 5.84$^{3.55}_{3.49}$ & 1.4$^{+0.6}_{-0.2}$ & 1 & 1,2 \\[5pt]
    Kepler-438 & 6.03$^{3.45}_{3.49}$ & 4.4$^{+0.8}_{-0.7}$ & 2 & 1,2 \\[5pt]
    KIC-7340288 & 5.77$^{3.50}_{3.31}$ & - & - & - \\[5pt]
    Kepler-441 & 4.99$^{2.04}_{1.13}$ & 1.9$^{+0.5}_{-0.4}$ & 2 & 1,2 \\[5pt]
    Kepler-442 & 6.09$^{3.41}_{3.47}$ & 2.9$^{+8.1}_{-0.2}$ & 2 & 1,2 \\[5pt]
    HD 40307 & 4.65$^{3.63}_{2.79}$ & 6.9 $\pm$ 4.0 & 3 & 1 \\[5pt]
    Kepler-62 & 5.82$^{3.55}_{3.48}$ & 4.0 $\pm$ 0.6 & 2 & 1,2 \\[5pt]
    Kepler-1544 & 3.50$^{2.30}_{1.12}$ & 3.90$^{+7.30}_{-0.80}$ & 1 & 2 \\[5pt]
    Kepler-452 & 4.62$^{2.67}_{1.36}$ & 6 $\pm$ 2 & 4 & 2 \\[5pt]
    \enddata
    \tablenotetext{a}{References: 1.\citet{2017AJ....154..264T} 2.\citet{2015ApJ...800...99T} 3.\citet{2016AA...585A...5B} 4.\citet{2015AJ....150...56J}}
    \tablenotetext{b}{Measurement Techniques: 1.Gyrochronology 2.Isochrone}
\end{deluxetable}

Since the age determination is easily automated, we have determined ages for a sample of TESS continuous viewing zone (CVZ) targets from the TESS Input Catalog (TIC). The sample of 2768 stars was retrieved from the TIC version 8.1 via the Mikulski Archive for Space Telescopes (MAST) at the Space Telescope Science Institute. The sample spans masses of 0.5-1.3 M$_\odot$, ecliptic latitudes of $\theta <-78^\circ$ and $\theta >78^\circ$, TESS apparent magnitudes $T_{mag}<10$, and all stars have luminosities consistent with dwarf stars. Potential planets in these systems will receive the most observation time from TESS and likely from both ground- and space-based follow-up missions. Stellar ages will be essential for placing probabilistic constraints on the habitability potential of each for future target prioritization. A sample of estimated ages is included in Table \ref{tab:tess_ages}, which includes the stellar parameters taken from the TIC.

\begin{deluxetable}{cccccc}[htb!]
    \tablecolumns{6}
    \tabletypesize{\scriptsize}
    \tablewidth{0pt}
    \tablecaption{TESS CVZ Stellar Ages\label{tab:tess_ages}}
    \tablehead{\colhead{TIC} & \colhead{Age$_{Tycho}$ [Gy]} & \colhead{$M/M_\odot$} & \colhead{$T_{eff}$ [K]} & \colhead{$L/L_\odot$} & \colhead{[M/H]}}
    \startdata
    55454149  &	6.62$^{1.59}_{2.32}$    & 1.03 $\pm$ 0.13 & 5754 $\pm$	132 &	2.45 $\pm$ 0.08 & ... \\ [5pt]
    141912469 &	2.79$^{1.83}_{1.45}$     & 1.20 $\pm$ 0.18 & 6232 $\pm$	138 &	2.68 $\pm$ 0.12 & -0.59 $\pm$ 0.09 \\ [5pt]
    38844604  &	5.67$^{2.29}_{2.48}$   & 1.05 $\pm$ 0.13 & 5813 $\pm$	131 &	3.27 $\pm$ 0.12 & ... \\ [5pt]
    350824257 &	6.16$^{1.65}_{1.71}$   & 1.05 $\pm$ 0.13 & 5830 $\pm$	117 &	2.76 $\pm$ 0.10 & -0.10 $\pm$ 0.10 \\ [5pt]
    233080190 &	3.05$^{0.92}_{0.51}$   & 1.23 $\pm$ 0.18 & 6290 $\pm$	128 &	4.45 $\pm$ 0.16 & ... \\ [5pt]
    33879314  &	5.06$^{1.95}_{3.17}$     & 1.04 $\pm$ 0.12 & 5782 $\pm$	104 &	1.26 $\pm$ 0.05 & -0.12 $\pm$ 0.05 \\ [5pt]
    280162266 &	2.85$^{1.50}_{1.05}$     & 1.24 $\pm$ 0.19 & 6304 $\pm$	129 &	2.85 $\pm$ 0.12 & ... \\ [5pt]
    289540757 &	2.04$^{3.80}_{0.95}$     & 1.12 $\pm$ 0.15 & 6034 $\pm$	119 &	1.22 $\pm$ 0.04 & ... \\ [5pt]
    441724181 &	5.72$^{2.29}_{2.59}$     & 1.06 $\pm$ 0.13 & 5865 $\pm$	124 &	1.52 $\pm$ 0.05 & ... \\ [5pt]
    232629681 &	4.80$^{2.64}_{1.70}$ & 1.10 $\pm$ 0.14 & 5987 $\pm$	108 &	4.81 $\pm$ 0.17 & -0.04 $\pm$ 0.03 \\ [5pt]
    55295030  &	6.24$^{1.76}_{3.09}$     & 1.11 $\pm$ 0.14 & 6005 $\pm$	119 &	2.55 $\pm$ 0.10 & -0.53 $\pm$ 0.05 \\ [5pt]
    220411843 &	4.18$^{4.23}_{2.91}$     & 0.94 $\pm$ 0.12 & 5398 $\pm$	140 &	0.58 $\pm$ 0.02 & 0.06 $\pm$ 0.05 \\ [5pt]
    219898046 &	6.41$^{1.95}_{2.35}$    & 1.01 $\pm$ 0.13 & 5697 $\pm$	131 &	2.52 $\pm$ 0.07 & ... \\ [5pt]
    149625812 &	8.12$^{1.35}_{2.35}$     & 1.03 $\pm$ 0.13 & 5751 $\pm$	128 &	1.87 $\pm$ 0.07 & ... \\ [5pt]
    287140180 &	5.62$^{2.32}_{2.44}$   & 1.04 $\pm$ 0.12 & 5778 $\pm$	112 &	3.31 $\pm$ 0.09 & ... \\ [5pt]
    198161860 &	2.41$^{2.42}_{1.37}$    & 1.25 $\pm$ 0.19 & 6330 $\pm$	127 &	2.00 $\pm$ 0.07 & ... \\ [5pt]
    441812317 &	3.52$^{2.20}_{1.25}$     & 1.16 $\pm$ 0.16 & 6124 $\pm$	124 &	2.93 $\pm$ 0.10 & ... \\ [5pt]
    256299260 &	3.14$^{1.22}_{0.32}$   & 1.20 $\pm$ 0.17 & 6219 $\pm$	119 &	4.33 $\pm$ 0.18 & ... \\ [5pt]
    289572073 &	4.80$^{0.70}_{1.12}$   & 1.12 $\pm$ 0.14 & 6024 $\pm$	120 &	3.75 $\pm$ 0.14 & ... \\ [5pt]
    \enddata
    \tablecomments{Table \ref{tab:tess_ages} is published in its entirety in the machine-readable format. A portion is shown here for guidance regarding its form and content.}
\end{deluxetable}

\begin{figure}[htb!]
    \centering
    \begin{tabular}{cc}
    \includegraphics[width=0.47\textwidth]{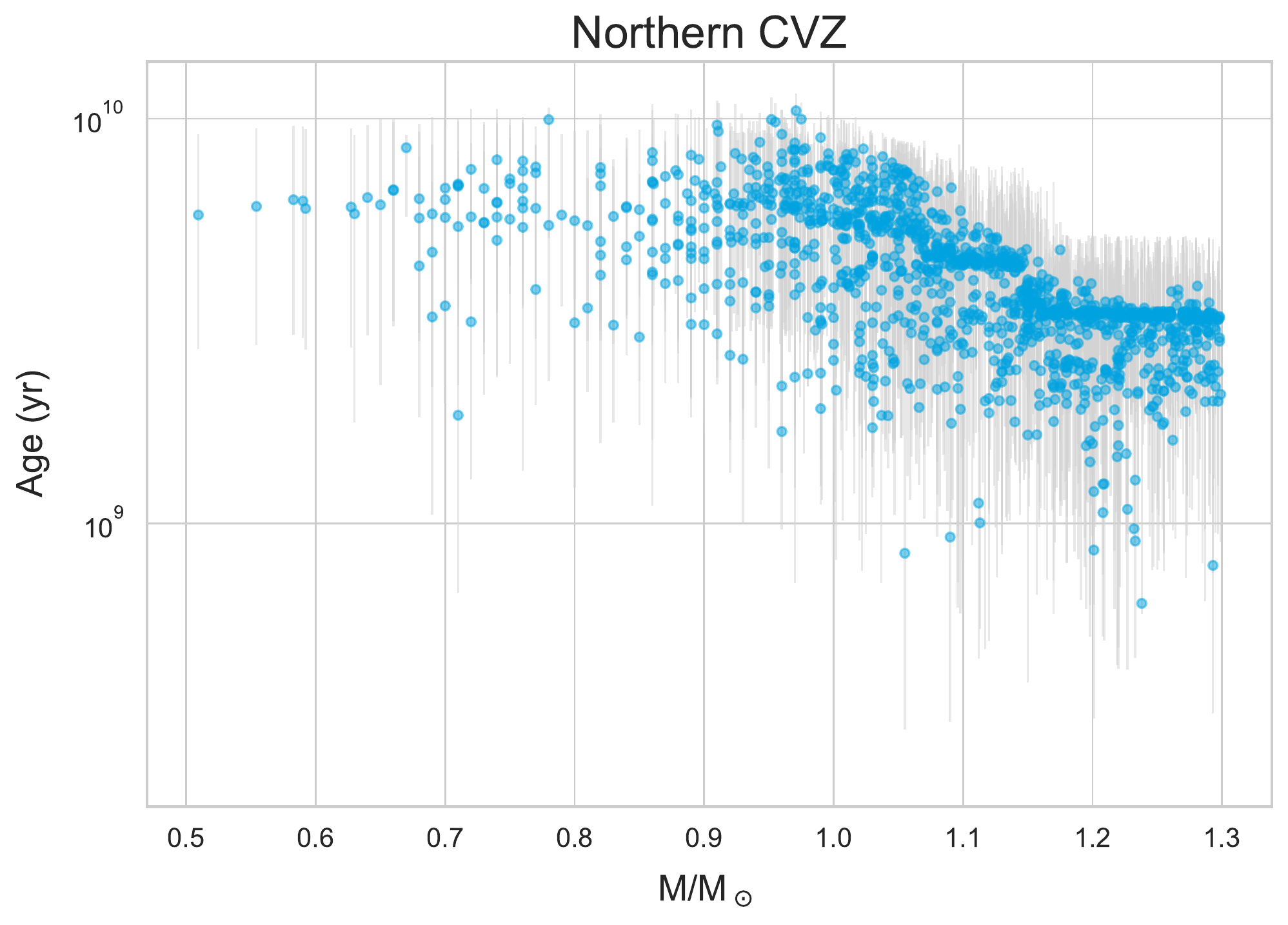} &
    \includegraphics[width=0.47\textwidth]{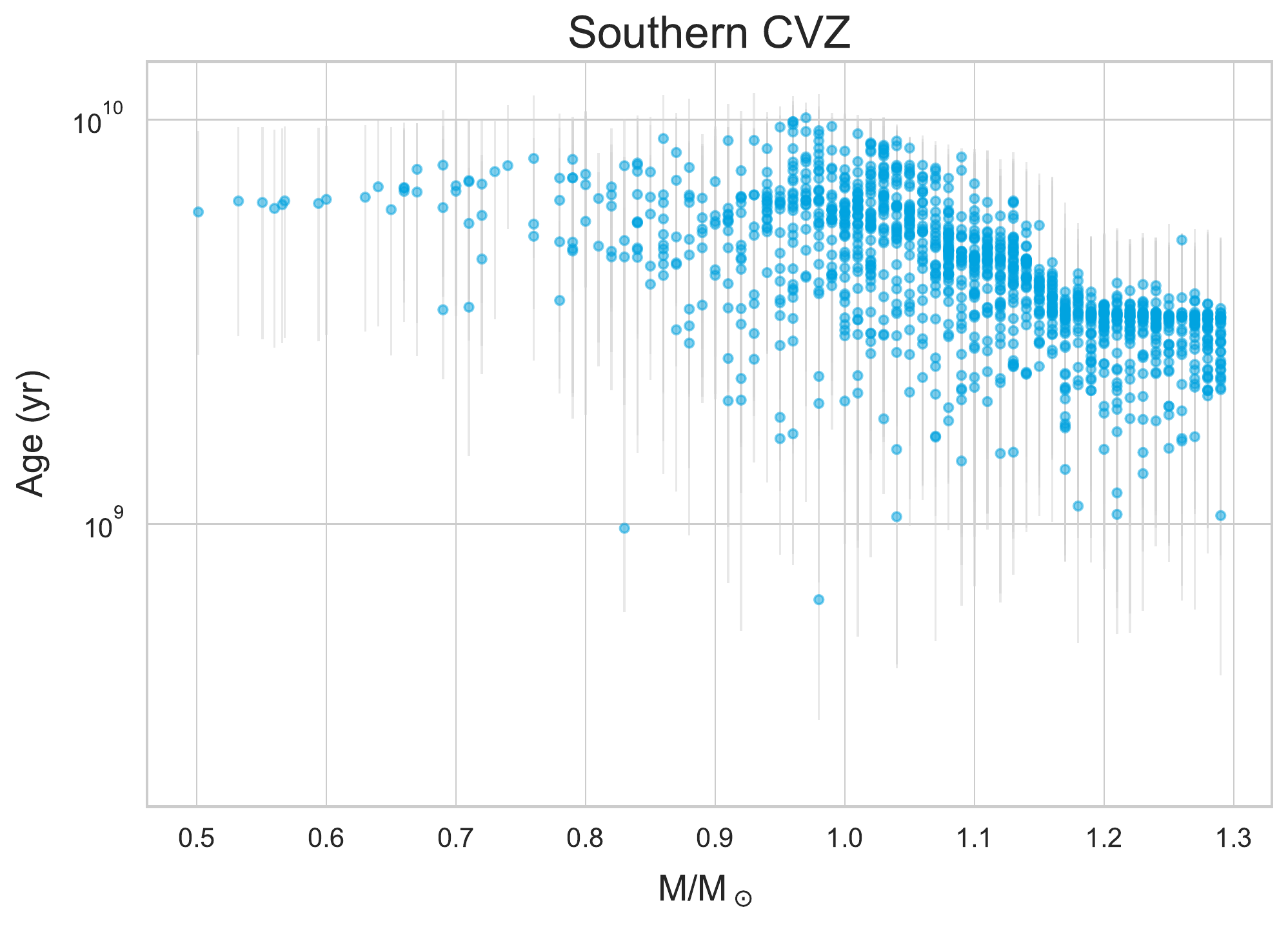}
    \end{tabular}
    \caption{Tycho best-fit model ages for TESS Northern and Southern CVZ sample of F, G, K, and early-M stars, using stellar properties from the TIC version 8.}
    \label{fig:tic_ages}
\end{figure}

An overview of the results of this age analysis is shown in Figure \ref{fig:tic_ages}. Notably, we see a concentrated band stars at higher ages and a more dispersed group of stars at lower ages. This feature is present in both the Northern and Southern CVZ samples, indicating no directional correlation. In addition, only one star in the sample, TIC 30270183, is known to be a member of a cluster or moving group. \citet{2015ApJ...808..132H} observed two distinct populations in compositional ([$\alpha$/Fe] versus [Fe/H]) space at $R>5$ kpc in the Milky Way's disk. One population is roughly solar-$\alpha$ and spans a large range of [Fe/H], which merges with a lower-$\alpha$ population at super-solar [Fe/H]. This could be indicative of a lower metallicity, older stellar population and a higher metallicity, younger population. Determining the exact nature of these age bands is beyond the scope of this work, but we will further investigate this in the future.

\section{CHZ$_2$ Planet Profiles}\label{sec:chz2_prof}

\begin{figure}[htb!]
        \centering
        P(CHZ$_2\mid$Z,M,A) vs. Distance from Star (AU)
        \begin{tabular}{@{}c@{}c@{}c@{}}
        \includegraphics[width=0.33\textwidth]{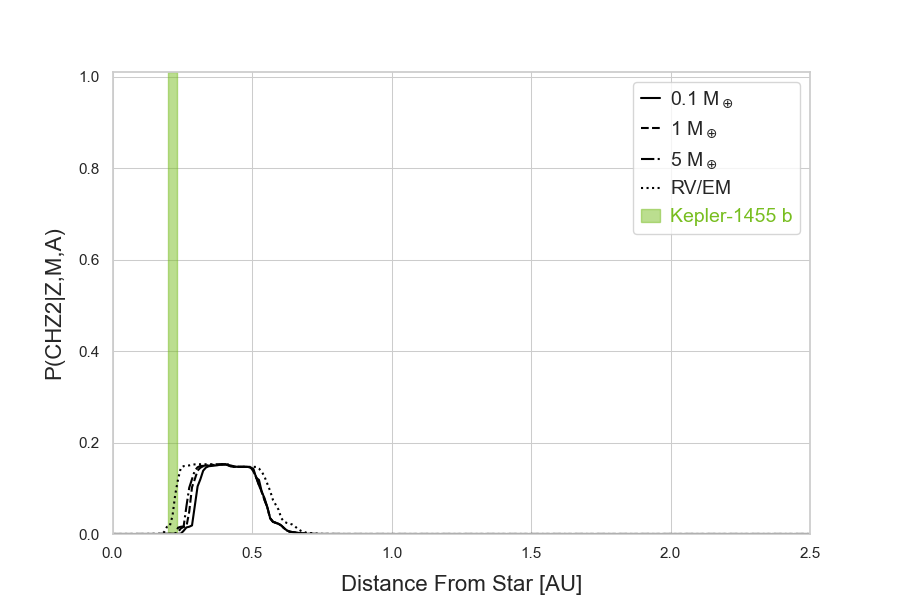} &
        \includegraphics[width=0.33\textwidth]{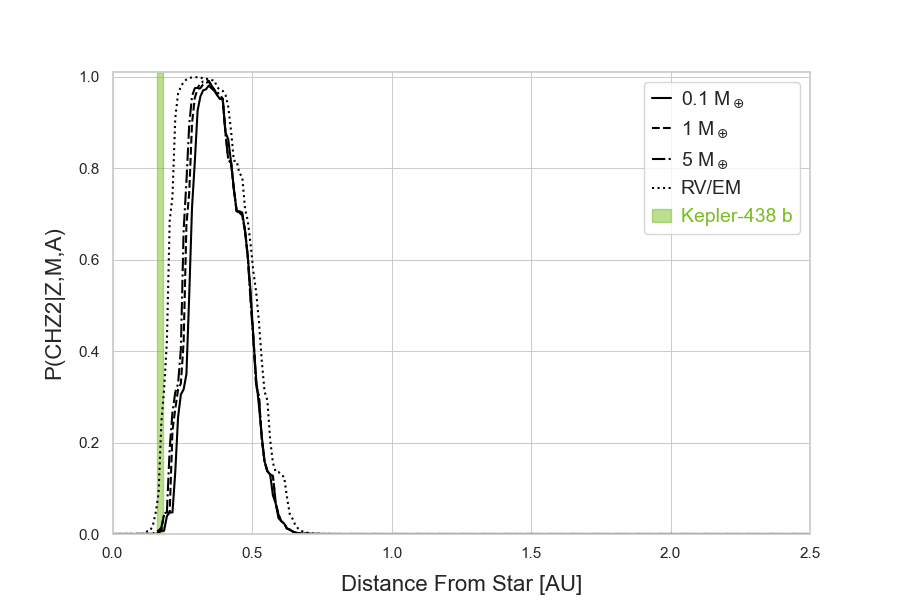} &
        \includegraphics[width=0.33\textwidth]{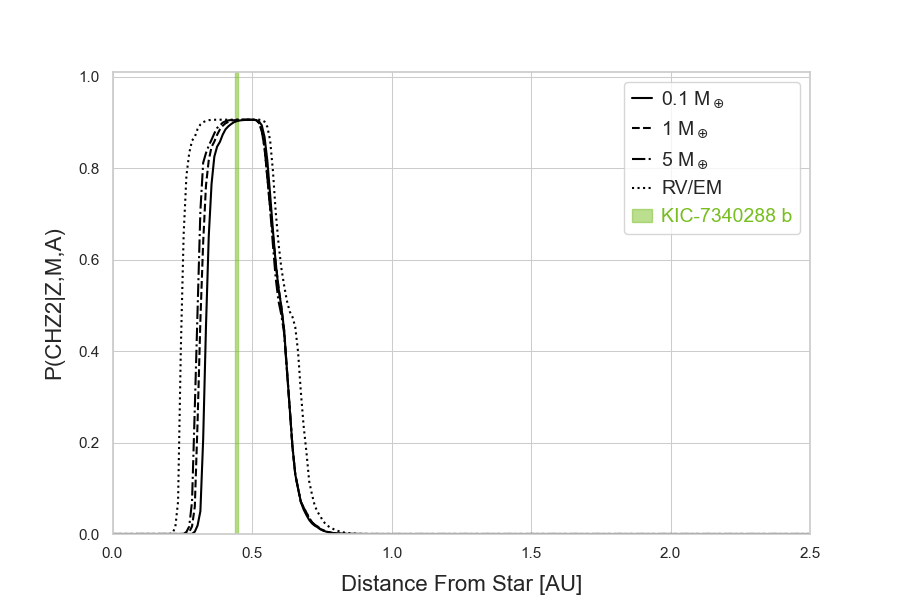} \\
        \includegraphics[width=0.33\textwidth]{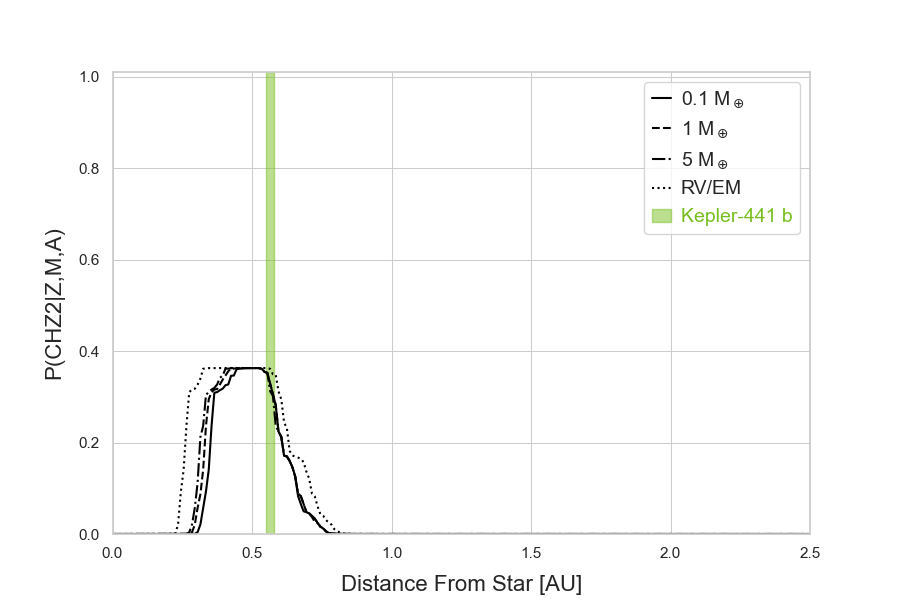} &
        \includegraphics[width=0.33\textwidth]{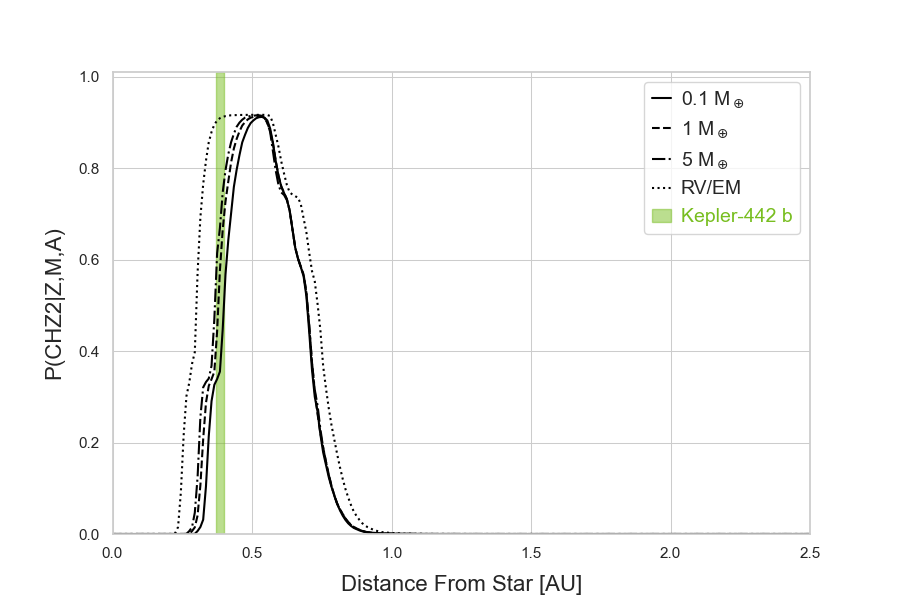} &
        \includegraphics[width=0.33\textwidth]{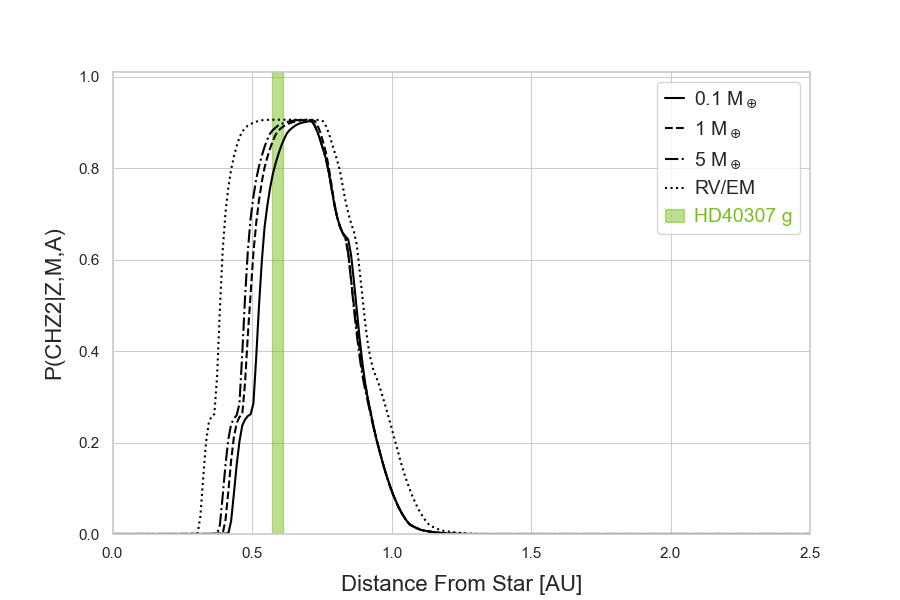} \\
        \includegraphics[width=0.33\textwidth]{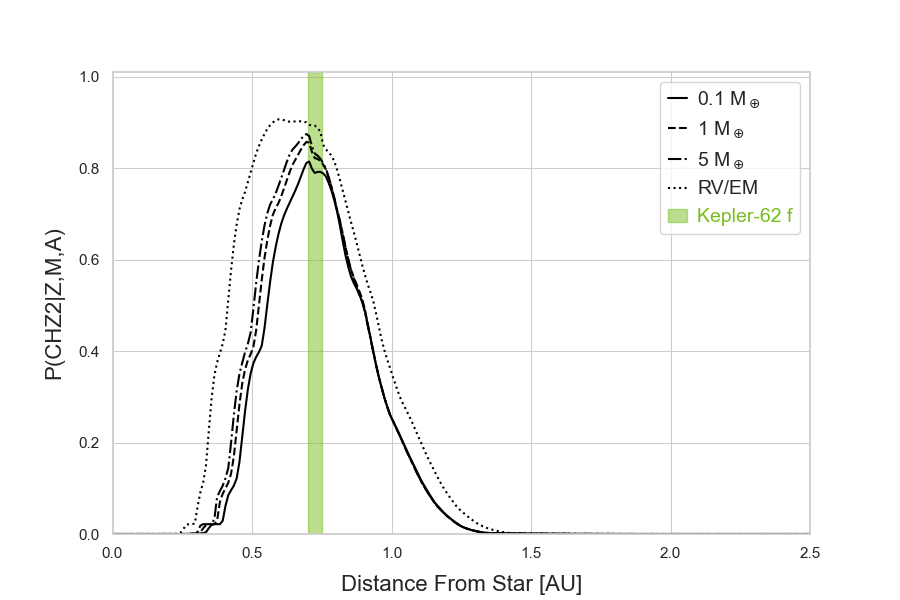} &
        \includegraphics[width=0.33\textwidth]{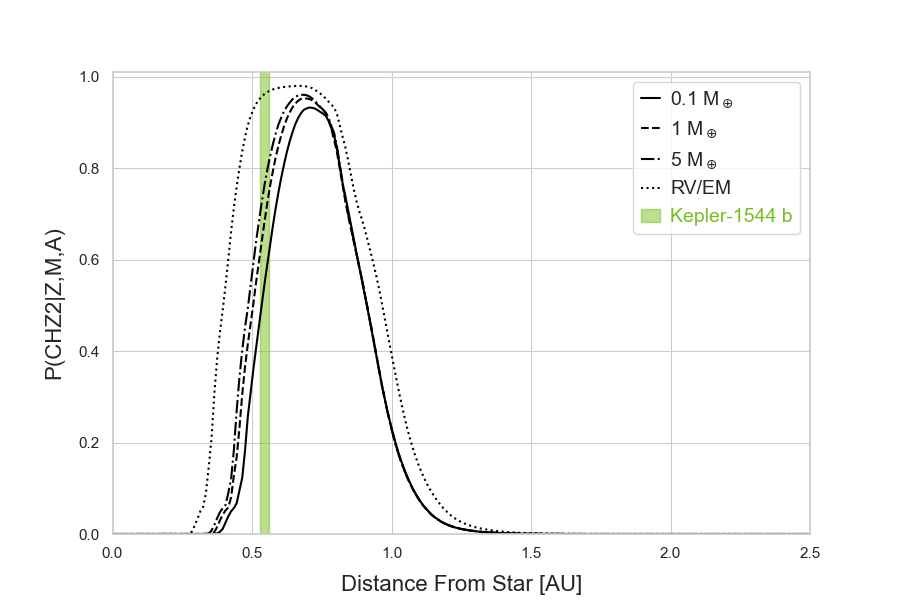} &
        \includegraphics[width=0.33\textwidth]{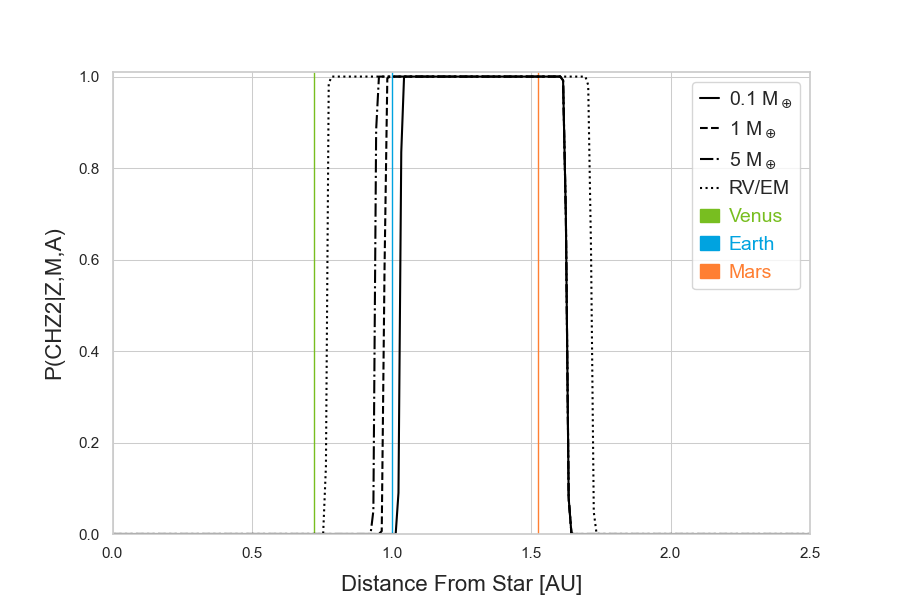} \\
        & \includegraphics[width=0.33\textwidth]{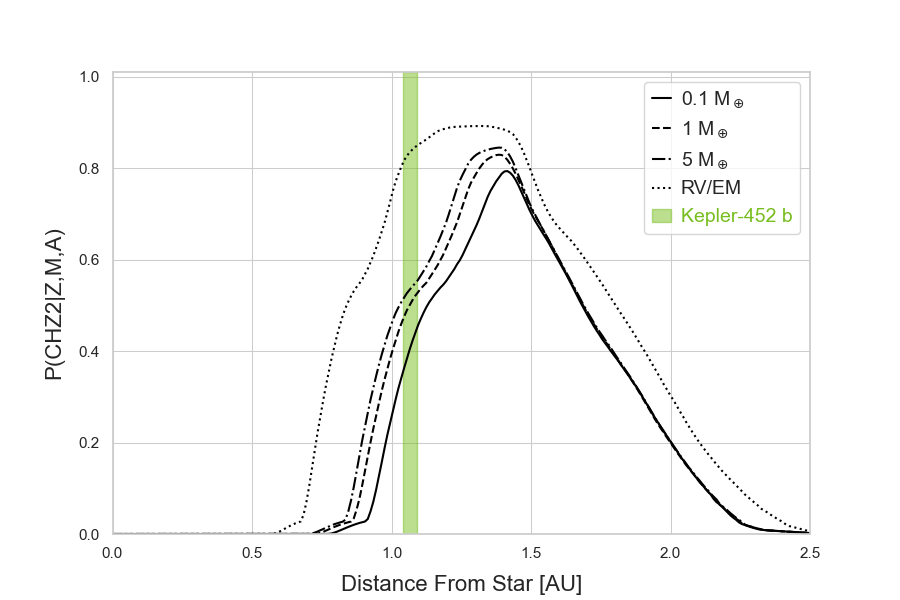} &
        \end{tabular}
        \caption{P(CHZ$_2\mid$Z,M,A) for Kepler-1455 b, Kepler-438 b, KIC-7340288 b, Kepler-441 b, Kepler-442 b, HD 40307 g, Kepler-62 f, Kepler-1544 b, the Sun (Venus, Earth, Mars), and Kepler-452 b assuming a 0.1, 1, and 5 M$_\oplus$ runaway greenhouse conservative inner habitable zone (IHZ) and maximum greenhouse conservative outer habitable zone (OHZ), as well as an optimistic case assuming a Recent Venus (RV) IHZ and Early Mars (EM) OHZ. The orbital range of planets are indicated by filled rectangles.}
        \label{fig:res1}
\end{figure}

CHZ$_2$ posterior probabilities for the sample of 9 potentially habitable exoplanets, as well as Venus, Earth, and Mars, are included in Table \ref{tab:prob}, with the full CHZ$_2$ distributions for each star, including the Sun, displayed in Figure \ref{fig:res1}. $<P_{0.1}>$, $<P_{1}>$, $<P_{5}>$, and $<P_{RV/EM}>$ are the posterior probabilities averaged over the orbital range of the planet for each HZ model from Section \ref{sec:hz_mod} (0.1 $M_\oplus$, 1 $M_\oplus$, 5 $M_\oplus$, and Recent Venus/Early Mars).

\begin{deluxetable}{ccccccccc}[htb!]
        \tablecolumns{9}
        \tabletypesize{\scriptsize}
        \tablewidth{0pt}
        \tablecaption{Planetary CHZ$_2$ Posterior Likelihoods\label{tab:prob}}
        \tablehead{\colhead{Planets} & \colhead{M$_\oplus$} & \colhead{R$_\oplus$} & \colhead{Orbit [au]} & \colhead{$<P_{0.1}>$} & \colhead{$<P_{1}>$} & \colhead{$<P_{5}>$} & \colhead{$<P_{RV/EM}>$}}
        \startdata
        Kepler-1455 b & ... & 1.75 & 0.20-0.23 & 0.0 & 0.0 & 0.0 & 0.048 \\
        Kepler-438 b & ... & 1.12 & 0.16-0.18 & 0.001 & 0.005 & 0.007 & 0.068 \\
        KIC-7340288 b & ... & 1.51 & 0.44-0.45 & 0.900 & 0.905 & 0.905 & 0.906 \\
        Kepler-441 b & ... & 1.462 & 0.55-0.58 & 0.345 & 0.345 & 0.34 & 0.362 \\
        Kepler-442 b & ... & 1.34 & 0.37-0.40 & 0.340 & 0.458 & 0.584 & 0.903 \\
        HD 40307 g & 7.1 & ... & 0.57-0.61 & 0.797 & 0.865 & 0.885 & 0.906 \\
        Kepler-62 f & ... & 1.531 & 0.70-0.75 & 0.801 & 0.838 & 0.850 & 0.894 \\
        Kepler-1544 b & ... & 1.685 & 0.53-0.56 & 0.523 & 0.661 & 0.743 & 0.959 \\
        Venus & 0.815 & 0.950 & 0.72 & 0.0 & 0.0 & 0.0 & 0.0 \\
        Earth & 1.00 & 1.00 & 1.00 & 0.0 & 1 & 1 & 1 \\
        Mars & 0.107 & 0.531 & 1.52 & 1 & 1 & 1 & 1 \\
        Kepler-452 b & ... & 1.511 & 1.04-1.09 & 0.402 & 0.499 & 0.533 & 0.833 \\
        \enddata
        \tablenotetext{}{\textbf{Note:} The orbital radius ranges indicated here are representative of $\pm 1\sigma$ semimajor axis orbits and these are calculated using the periods in Table \ref{tab:samp} and stellar masses in Table \ref{tab:hab_params}.}
\end{deluxetable}

Although all 9 exoplanets fall within the current HZ around their host star, we determine Kepler-1455 b and Kepler-438 b to have approximately P(CHZ$_2$) $\approx$ 0 for all cases, or essentially little to no chance of having been continuously in the HZ for 2 Gy. This indicates that they are situated too close to the inner edge to spend a significant amount of time in the HZ and will likely soon leave the HZ.

KIC-7340288 b, an unconfirmed super-Earth planet candidate \citep{2020AJ....159..124K}, is consistently a high probability target in our sample. The planet candidate's orbital range puts it at the peak CHZ$_2$ probability (P $\approx$ 0.9) for all HZ cases considered. Given KIC-7340288 b's relatively large size ($R=1.511 R_\oplus$), the planet candidate is more likely to have retained its atmosphere and would be better able to regulate the surface temperature and retain water. Therefore, our prediction that KIC-7340288 b has remained in a stable HZ environment for 2 Gy should be seen as a substantial indicator of the planet candidates potential habitability.

Notably, Earth resides at the inner edge of the CHZ$_2$ for the 1 $M_\oplus$ HZ model and is outside the inner edge for the 0.1 $M_\oplus$ model. Being that Earth is the only habitable planet we know of in the Universe, this shows that too conservative of HZ models are likely to exclude planets that have a high potential of being habitable. Although Mars has a 100\% probability of being in the CHZ$_2$ across all models and would generally be considered uninhabitable, it is better to include a potential Mars-like planet rather than excluding an Earth-like planet.

\section{Conclusions}\label{sec:conc}

This work builds upon \citet{2020AJ....159...55T} by further expanding the Tycho model space and adding a stellar age prior. The additions improve the framework's ability to estimate the time-dependent habitability of a planet given only limited knowledge of stellar properties and planetary orbital radius. We further applied this Bayesian method to analyzing the long-term habitability of 9 likely-rocky exoplanets with a high probability of being in the HZ (Table \ref{tab:prob}). The posterior probability distributions used priors of measured stellar metallicity, mass, and age fitted to Tycho stellar evolution models and HZ definitions from \citet{2013ApJ...765..131K,2014ApJ...787L..29K}. Two such exoplanets, Kepler-1455 b and Kepler-438 b, are shown to be unlikely to have spent 2 Gy in the HZ for this model. KIC-7340288 b, a recently discovered super-Earth planet candidate included in our sample, is consistently found to have the highest probability of having been in the HZ for 2 Gy.

The addition of an age prior and a method for estimating stellar ages from Tycho models will prove invaluable in future work estimating the continuous habitability of unstudied TESS candidates. By attaining fits to stellar ages comparable to published gyrochronology measurements and other isochrone ages, we demonstrate the ability to estimate ages for future potentially habitable systems with existing observations. We applied this method for age estimation to a sample of F, G, K, and early-M TESS CVZ stars from the TIC and produced ages for 2768 stars. Stars in the TESS CVZs will receive the most observing time from TESS and are the only TESS targets likely to yield detections of potentially habitable planets around Sun-like stars. The inner 5$^\circ$ of the TESS CVZs are also coincident with that of JWST, so many candidates found here are the most likely for JWST follow-up. The techniques presented here can be rapidly applied to candidates detected around HabEx and LUVOIR target stars.

In the near-term, we will include stellar evolution models down to 0.1 M$_\odot$, as this will enable the analysis of the remainder of potentially habitable systems, as well as the majority of the systems likely to be found with TESS. Increasing the resolution of the models in the lower-mass regime will provide better fits to K and M dwarfs as well. 

Our finding that Mars is given an equal probability to Earth of being in the CHZ$_2$ by our model is a prime example of the need for additional observational and model constraints in order to more accurately predict potential habitability. To explore the effects of introducing additional priors to the Bayesian method, we will include stellar and planetary properties important to stellar and planetary evolution, such as stellar [O/Fe] ratios, planetary composition, and stellar activity, an increasingly important factor in the study of M-dwarf systems. Expanding the suite of HZ models will provide additional confidence constraints on individual systems. Planetary mass, atmospheric composition, stellar multiplicity, and various other factors will impact the ability of liquid water to remain present on a planet for long enough for the presence of life to be remotely detectable. These properties can be added individually as priors as measurements and theoretical predictions become available.

\section{Acknowledgements}

The results reported herein benefitted from collaborations and/or information exchange within NASA’s Nexus for Exoplanet System Science (NExSS) research coordination network sponsored by NASA’s Science Mission Directorate. We would like to thank Eric Mamajek for information regarding spectral classification and current habitable zone exoplanets. We would also like to thank the reviewers for comments that lead to significant improvements in the quality of this work.


\bibliography{2_bib}{}
\bibliographystyle{aasjournal}



\end{document}